# Influence of the Graft Length on Nanocomposite Structure and Interfacial Dynamics

Anne-Caroline Genix,[1,*] Vera Bocharova,[2] Bobby Carroll,[3] Philippe Dieudonné-George,[1] Edouard Chauveau,[1] Alexei P. Sokolov,[2,3,4] and Julian Oberdisse [1,*]

[1]Laboratoire Charles Coulomb (L2C), Université de Montpellier, CNRS, F-34095 Montpellier, France

[2]Chemical Sciences Division, Oak Ridge National Laboratory, Oak Ridge, TN 37831, USA

[3]Department of Physics, University of Tennessee, Knoxville, TN 37996, USA

[4]Department of Chemistry, University of Tennessee, Knoxville, TN 37996, USA

* Corresponding authors: anne-caroline.genix@umontpellier.fr; julian.oberdisse@umontpellier.fr

**Abstract:** Both the dispersion state of nanoparticles (NPs) within polymer nanocomposites (PNCs) and the dynamical state of the polymer altered by the presence of the NP/polymer interfaces have a strong impact on the macroscopic properties of PNCs. In particular mechanical properties are strongly affected by percolation of hard phases, which may be NP networks, or dynamically modified polymer regions, or combinations of both. In this article, the impact on dispersion and dynamics of surface modification of the NPs by short monomethoxysilanes with eight carbons in the alkyl part ($C_8$) is studied. As a function of grafting density and particle content, polymer dynamics is followed by broadband dielectric spectroscopy and analyzed by an interfacial layer model, while the particle dispersion is investigated by small-angle X-ray scattering and analyzed by reverse Monte Carlo simulations. NP dispersion are found to be destabilized only at the highest grafting. The interfacial layer formalism allows to clearly identify the volume fraction of interfacial polymer, with its characteristic time. The strongest dynamical slow-down in the polymer is found for unmodified NPs, while grafting weakens this effect progressively. The combination of all three techniques enables a unique measurement of the true thickness of the interfacial layer, which is ca. 5 nm. Finally, the comparison between longer ($C_{18}$) and shorter ($C_8$) grafts provides unprecedented insight into the efficacy and tunability of surface modification. It is shown that $C_8$-grafting allows for a more progressive tuning, which goes beyond a pure mass effect.

# 1. Introduction

The versatility of the (mechanical, electrical, etc.) properties of polymer nanocomposites (PNCs) depends primarily on the system under study, namely the type, properties, and crosslinking of the polymer, and the size, surface chemistry and dispersion of the filler nanoparticles (NPs). [1-4] Different chemistries induce different particle interactions, and contribute thus to forming different NP dispersions, which may range from individually dispersed to networks, with obvious consequences on the mechanical properties. [5-9] For a given system, it is possible to tune these NP interactions by surface modification using small silane molecules. [10,11] The latter impact not only the interparticle potentials, but also the polymer-particle interactions. These interactions may include modification of van der Waals interactions, [12] or effects on the monomer cage and elastic barriers governing the segmental dynamics due to the change of surface nature (and roughness) from hard to soft surfaces. [13-15] The presence of the coating of the silica NPs introduces a different physical chemistry in terms of dielectric strength, hydrophobicity, rugosity and hardness, the latter affecting dynamical caging constraints. In this article, we therefore refer to the sum of the possible effects of the surface modification on polymer-particle interaction as "screening", as we cannot disentangle them in our experiments. Depending on these interactions, the thermodynamic properties of the polymer may change over some nanometric distance from the particle, and thus form an interfacial polymer layer. [16] The properties of this interfacial layer, which possesses a slowed-down dynamics and enhanced modulus in PNCs with attractive polymer-particle interactions, [17] in turn affect the mechanical properties of the entire sample, in particular if the interfacial layers connect and percolate. [18] It is thus both technologically relevant and fundamentally challenging to study the impact of silane surface modification on both particle dispersion and polymer dynamics simultaneously, and analyze the results in a combined approach. More specifically, it is the goal of the present article to explore the effect of silane chain length on both the structure and the interfacial layer dynamics, as well as interfacial regions overlap which may lead to the formation of percolated structures.

Polymer dynamics can be studied by quasi-elastic neutron scattering methods, with spatial and time resolution. On the other hand, broadband dielectric spectroscopy (BDS) is a laboratory method giving access to spatially unresolved dynamics over a large frequency range. [19] Due to its relative simplicity, it has been applied to many different nanocomposite systems, as long as some ion or dipole relaxation of relevance can be followed. Although we focus in the present analysis on the segmental ($\alpha$) relaxation and its shift in frequency with the system formulation, it is important to quantitatively include the neighboring processes in the description. Conductivity and Maxwell-Wagner-Sillars (MWS) polarization processes, which we have described in detail in the past, [20,21] are usually rather dominant on the low-frequency side of the BDS spectra. The subtle shift to lower frequencies of the $\alpha$-relaxation can thus only be determined if these contributions are weak and far from the $\alpha$-relaxation, or are modelled quantitatively. The $\beta$-relaxation at higher frequencies has also a measurable impact on the shape of the segmental relaxation. Thus, both need to be included in the models. Finally, in presence of a different (slowed-down) polymer phase around the nanoparticles, the non-additive interfacial layer model (ILM) [22] can be applied to describe the simultaneous presence of both bulk and interfacial segmental dynamics. [17,23] This interfacial layer model description is based on an average relaxation time over some polymer volume in contact with the surface. It is known from theoretical and simulation approaches [15] that the relaxation time follows a steep, double-exponential gradient in space, typically over some 5-10 monomer diameters. On the other hand, the present ILM analysis of the BDS data provides the interfacial layer volume fraction, and the characteristic time (distribution) of both phases. We anticipate here that the conversion from interfacial layer volume fraction to its thickness necessitates information on layer overlap and thus particle dispersion.

As mentioned above, the relative vicinity of NPs, including possible aggregation, has a strong impact on the macroscopic mechanical properties of the PNCs. Incidentally, particle arrangement in space also affects the overlap of dynamically slowed-down (i.e., high modulus) polymer interfacial layers, possibly contributing to the overall mechanical response. Particle dispersions can be measured by TEM, [24,25] with however limited statistical relevance of some picture of very small pieces of sample. In this respect, small-angle scattering, in particular of X-rays (SAXS), is a well-known technique to obtain average PNC structures in a quick and reliable way. [26] The price to pay is the difficulty of analysis, as particle correlations are intertwingled with particle size effects (polydispersity), and possible large-scale heterogeneities which are also included in the average. Traditional analysis of SAXS intensities is often based on reading off the position of a structure factor peak, if it exists, which is hoped to correspond to the most probable interparticle center-to-center distance encountered in the sample. [17,27] Much of the other information, may it be peak shape or low-q upturns, are often disregarded, although sometimes sophisticated theoretical integral-equation approaches, in particular PRISM, [28] may be used to predict partial structure factors of colloid-particle mixtures from thermodynamic interactions. [29,30] In the past, we have developed numerical methods based on a reverse Monte Carlo [31,32] approach. [33-35] Particle polydispersity is fully taken into account, and as a result sets of representative particle dispersions compatible with the experimentally observed scattering are obtained, in a simulation box of roughly micrometric size. It is then straightforward to analyze these dispersions statistically, in particular in terms of particle spacing or aggregation. One may note that in the past we have started to combine TEM with SAXS analyses of PNCs. [36,37]

In a recent article, we had studied the effect of grafting of alkyl monomethoxysilane with 18 carbon atoms in the alkyl part ($C_{18}$) on interfacial polymer dynamics and particle dispersions, by using BDS and SAXS. [12] In the present article, the effect of a shorter $C_8$-silane is studied using the same methods and analysis, varying both the grafting density and the particle volume fraction. By comparing to the previously studied $C_{18}$-system, the impact of the alkyl chain length is highlighted. In a second time, a recent methodological approach is further developed. Reverse Monte Carlo simulations provide sequences of particle configurations of scattering compatible with the observed SAXS signals. This has been combined with the BDS results, which provide the volume fraction of interfacial layers. The result is a more realistic estimation of the interfacial layer thickness, because it avoids the idealized vision of perfectly dispersed particles underlying simple IPS equations, [38] while our approach allows taking into account polydispersity and layer overlap caused by NPs in close vicinity. This combined method is now investigated further by analyzing the evolution of the interfacial layer thickness with its volume fraction for each sample, i.e., for different types of experimentally observed dispersions.

## 2. Materials and Methods

**Nanoparticles and surface modification:** The silica NPs were synthesized in ethanol by a modified Stöber method with the final NP concentration of 16 mg/mL. For the functionalization step, the NP suspension was used as is without further purification. It was characterized by SAXS at high dilution (0.3%v). The scattered intensity is shown in the SI. It revealed a log-normal size distribution of spheres ($R_0$ = 8.4 nm, σ = 18%), leading to an average NP radius of R = 8.5 nm.

Surface modification of the NPs was performed with *n*-octyldimethylmethoxysilane ($CH_3(CH_2)_7Si(CH_3)_2OCH_3$, termed $C_8$) from Gelest. The grafting reaction was conducted at 323 K for 3 days in ethanol. To achieve different grafting densities ranging from zero (bare NPs) to ca. 3 nm$^{-2}$, different amounts of silanes were added to the NP suspension: 50 mL of the suspension were mixed with 120, 750, 1200, and 1500 μL of $C_8$ silane. After the reactions were completed, the surface-modified NP suspensions were dialyzed against ethanol for 3 days. The grafting densities were determined by thermogravimetric analysis (TGA, TA instrument, Discovery, 5 K/min under air) using the weight loss between 473 and 873 K corresponding to the thermal decomposition of the grafted silanes. [39] The TGA curves are given in SI (Figure S1). The resulting grafting density of $C_8$-molecules on the silica NPs, between 0.8 and 2.9 /nm², are given in Table 1.

**Nanocomposite formulation:** Four series in surface modification at nominal particle volume fractions of 2%v, 15%v, 20%v, and 30%v have been prepared. The polymer, poly(2-vinylpyridine) (P2VP) with a weight-average MW of 35.9 kg.mol$^{-1}$ (polydispersity index = 1.07), was purchased from Scientific Polymer Products Inc., and used as received. The radius of gyration of the chain is 5.2 nm. The polymer dissolved in ethanol (66 mg/mL) and (bare or surface-modified) NP suspension also in ethanol (16 mg/mL) were mixed for at least 12 hours, then filtered through a 200 nm Teflon filter. The final PNCs were formed by evaporating the solvent at room temperature followed by drying in a vacuum oven at 393 K for 2 days. All samples were hot-pressed at 423 K, and they were further annealed under vacuum at 393 K for 3 days before the SAXS and BDS measurements. The silica fractions in PNCs were obtained by TGA (20 K/min, under air) from the weight loss between 433 and 1073 K. The NP volume fractions, $\Phi_{NP}$, were determined by mass conservation using the density of neat P2VP ($\rho_{P2VP}$ = 1.19 g·cm$^{-3}$ by pycnometry) [17] and silica ($\rho_{NP}$ = 2.27 g·cm$^{-3}$ by SANS) [40]. They are reported in Table 1.

**Table 1.** NP volume fractions in PNCs, and $C_8$-grafting densities on NPs suspended in solvent, both determined by TGA.

|            | Bare  | $C_8$ 0.8/nm² | $C_8$ 1.3/nm² | $C_8$ 2.4/nm² | $C_8$ 2.9/nm² |
|------------|-------|---------------|---------------|---------------|---------------|
| 2%v-series | 2.0%  | 1.9%          | 1.9%          | 1.5%          | 0.3%          |
| 15%v-series| 15.3% | 13.3%         | 12.4%         | 11.9%         | 12.6%         |
| 20%v-series| 22.4% | 21.1%         | 19.5%         | 18.5%         | 19.0%         |
| 30%v-series| 30.7% | 28.1%         | 26.5% 26.1%   |               | 25.2%         |

**BDS:** BDS measurements were conducted on a broadband dielectric spectrometer (Novocontrol Alpha) and a Quatro Cryosystem temperature controller with a stability of ± 0.1 K. The complex dielectric permittivity, $\varepsilon^*(\omega) = \varepsilon'(\omega) - i\varepsilon''(\omega)$, was measured in the frequency range from $2.10^{-2}$ to $10^7$ Hz ($\omega = 2\pi f$) using disk-shaped samples with a diameter of 20 mm and a typical thickness of 0.15 mm. The samples (without spacer) were sandwiched between two gold-plated electrodes forming a capacitor. They were first annealed for 1 h at 433 K in the BDS cryostat under nitrogen flow to ensure that both the real and imaginary parts of the permittivity became constant in the probe frequency range. Then, isothermal frequency measurements were performed at 423 K, and from 303 K down to 233 K with an interval of 10 K to specifically follow the $\beta$-relaxation of P2VP. A measurement at the lowest measurable temperature of 103 K was performed to normalize the permittivity values. After that, the samples were measured again at 293 and 433 K to check reproducibility. The normalization procedure of PNCs is described in detail in [17] considering two-phase heterogeneous materials [19] with the high-frequency limit of the real part $\varepsilon_\infty$ = 3.05 and 3.9 for the polymer and silica, respectively. It allows getting rid of possible artifacts (mostly thickness variations) and leads to the dielectric spectra in absolute values.

**Dielectric analysis:** We aim at describing the segmental relaxation of P2VP in PNCs by fitting simultaneously the real and imaginary parts of the permittivity. As reported previously, the contributions of two phases are considered: the interfacial layer close to NP surfaces and the unmodified bulk polymer far from the particles. In this case, the contributions of each component are not additive, and the interference terms are explicitly taken into account in the interfacial layer model (ILM) for heterogeneous systems. [22] Detailed equations of the model are given in [23]. The bulk polymer contribution was described by a Havriliak-Negami (HN) function

$$\varepsilon^*(\omega) = \varepsilon_\infty + \frac{\Delta\varepsilon}{[1+(i\omega\tau_{HN})^\gamma]^\delta} \tag{1}$$

having the same spectral shape parameters γ and δ, dielectric strength Δε, and timescale $\tau_{HN}$ as the dielectric function of the neat polymer measured independently. The free parameters of the ILM are the volume fraction of interfacial layer, $\Phi_{IL}$, and its dielectric function, $\varepsilon_{IL}^*(\omega)$, which is well-described by a symmetrical HN-process (δ = 1). In the following, the relaxation times are defined by $\tau_{HN}$ related to the peak position in frequency $f_{max}$, which is used to determine the relaxation time $\tau_{max} = 1/(2\pi f_{max})$.

The β-process of P2VP was described in the low-T range (233 – 303 K), where it can be observed alone in the accessible frequency window, using a single symmetrical HN function. It was included in ILM using an extrapolation of the low-T data of each sample to account for the high-frequency contribution of the secondary dynamics to the α-process. Finally, a purely dissipative d.c. conductivity term and a Maxwell–Wagner–Sillars (MWS) process described by a symmetrical HN-process were also added systematically to describe the low-frequency part of the dielectric spectra of PNCs. The MWS process is associated with interfacial polarization effects in the presence of particles. [41]

**SAXS:** Small-angle X-ray measurements were performed with a wavelength λ = 1.54 Å (copper target) on an in-house setup of the Laboratoire Charles Coulomb, "Réseau X et gamma", University of Montpellier (France) was employed using a high-brightness low-power X-ray tube, coupled with aspheric multilayer optics (GeniX$^{3D}$ from Xenocs). It delivered an ultralow divergent beam (0.5 mrad). The scattered intensities were measured by a 2D "Pilatus" pixel detector at a single sample-to-detector distance D = 1900 mm, leading to a q range from $4\times10^{-3}$ to 0.2 Å$^{-1}$. The scattering cross-section per unit sample volume dΣ/dΩ (in cm$^{-1}$), which we term scattered intensity I(q), was obtained by using standard procedures including background subtraction and calibration. [42]

**Scattering analysis: reverse Monte Carlo (RMC).** The scattered intensity is described using a reverse Monte Carlo simulation, following previous approaches. [33-35] N polydisperse spherical particles are placed in a cubic simulation box with periodic boundary conditions, of dimension $L_{box} = 2\pi/q_{min}$, where $q_{min}$ is the experimental minimum q-value, such that the total volume fraction $\Phi_{NP}$ corresponds to the experimental one of the sample. The scattered intensity of the particles in the simulation box is calculated based on the individual size of each particle, and converted to absolute units using the contrast based on the scattering length densities of silica and the polymer: $\rho_{SiO2}$ = 19.49 10$^{10}$ cm$^{-2}$, $\rho_{P2VP}$ = 10.93 10$^{10}$ cm$^{-2}$. The calculation is split in the Debye formula [43] at high q, and a lattice calculation avoiding box contributions [35,44,45] at low q. Particles are then moved around randomly while following a simulated annealing procedure leading to agreement of the theoretically predicted apparent structure factor S(q) with the experimental one. In all cases, excluded volume of the particles is respected. Particle configurations that are compatible with the experimental intensity are saved regularly, and can be analyzed a posteriori, e.g., in terms of interparticle spacing. S(q) and any statistical measures are averaged over different particle configurations and represent the result of the simulation.

It is worth mentioning that depending on the region in q-space, experimental intensities and form factors vary by orders of magnitude. Calculating S(q) by division of two functions leads to high errors at large-q, which is why the highest q-values have generally been discarded in our analysis. TGA was systematically used to assess the silica content. A successful cross-check is shown in the SI, where the structures of two nominally identical PNCs produced at different moments are compared, and virtually perfect agreement is found.

## 3. Results and discussion

### A. Nanoparticles in solvent and polymer: dilute conditions

The shape of the bare and grafted NPs has been studied by SAXS in suspension in ethanol, under high dilution. The corresponding form factor is shown in the SI (Figure S2), and one can see that the superposition of the data is remarkable, across the entire q-range. This implies that although the grafting has taken place as proven by TGA, it has neither influence on the contrast in ethanol – i.e., the grafted and solvated layer is invisible to X-rays –, nor on the particle dispersion at this high dilution. The corresponding theoretical form factor is also superimposed to the data, and it corresponds to the log-normal size distribution of spheres reported in the methods section.

These particles have then been incorporated into a P2VP-matrix. A first series of samples at a nominal volume fraction of 2%v has been prepared and studied by SAXS. The resulting intensity curves are given in the SI, and they show that some modification of the scattering length density around the particles is visible in the polymer matrix. Although the observed deviations are quite small and not visible at low grafting density, they present a first structural evidence of the impact of grafting. At low q, the samples with intermediate grafting show a slight decrease, indicating a change in NP interactions towards short-range repulsion. The highest grafting density has a clearly different shape with a dip in intensity called a correlation hole [46], followed by a low-q increase, and it is a typical signature of aggregation in this nanocomposite.

The structure of PNCs at 2%v has also been studied by TEM. In Figure 1, the series in grafting density up to 2.4 /nm² is shown, and compared to a series at ten times higher silica content. These pictures illustrate that the NPs are rather well dispersed under all conditions, i.e., there are no large structural heterogeneities.

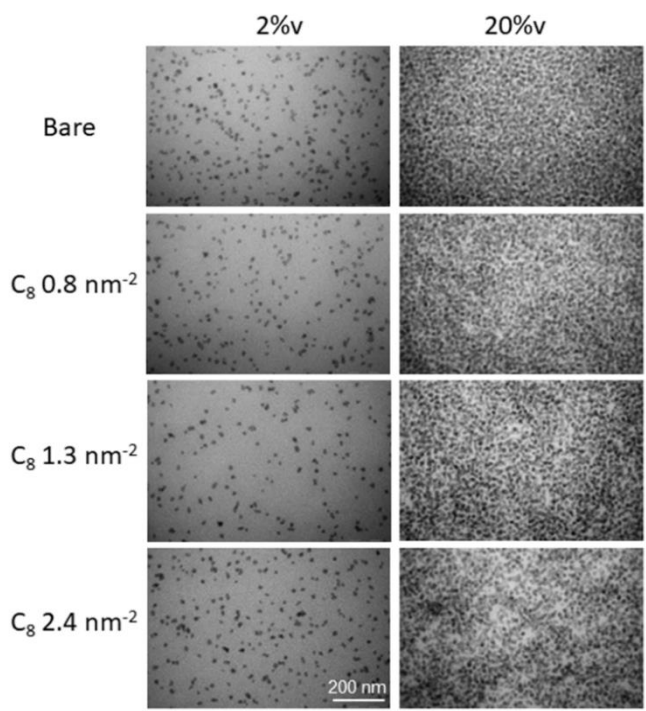

**Figure 1.** TEM micrographs of P2VP PNCs at nominal NP volume fraction of 2%v (**left**) and 20%v (**right**) with $C_8$-grafted silica. The grafting densities (0 to 2.4 nm$^{-2}$) vary from top to bottom as indicated in the legend. The exact volume fractions of all samples are given in Table 1.

At the higher volume fraction (ca. 20%v) in Figure 1, the dispersion of the particles is seen to remain rather homogeneous. At the highest grafting densities, the occurrence of more whitish silica-free zones indicates the presence of spatial fluctuations in particle density. This is impossible to see in the left-hand figure, due to the high dilution and thus important average particle distance. Scattering, however, is able of picking up such small fluctuations, as shown for the 2%v-samples in the SI. At 20%v, the spatial averaging as performed by the SAXS experiments will indeed confirm these heterogeneities which develop upon grafting.

**B. Dynamical properties of the NP-polymer interface probed by BDS**

Broadband dielectric spectroscopy has been used to probe the segmental α-relaxation associated with the polymer glass transition. In the dielectric spectra, the α-process is surrounded at lower frequencies by ionic conductivity and MWS-processes related to the filler nanoparticles, and at higher frequencies by the secondary β-relaxation of P2VP. All processes have been included in our analysis, in order to extract the segmental dynamics in a trustworthy way. In Figure 2, the dielectric loss spectra of the 30%v PNC series are presented (15 and 20%v can be found in the SI in Figure S3).

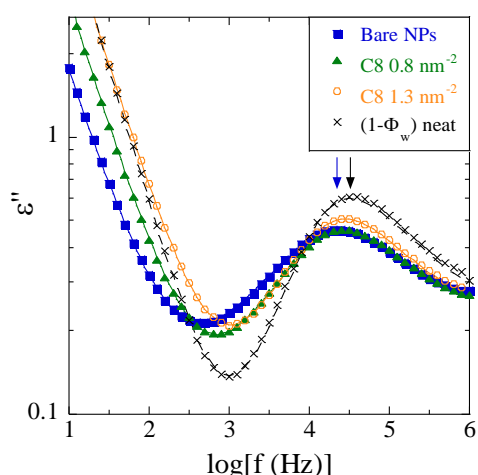

**Figure 2.** Comparison of the dielectric loss spectra of neat P2VP (black crosses, data normalized to the weight polymer fraction, 1 - $\Phi_w$) and PNCs with different surface modifications of the silica NPs for the series with nominal $\Phi_{NP}$ = 30%v at T = 423 K. Arrows indicate the position of the maximum of the loss peak in neat polymer (black) and PNC with bare NPs (blue). The black line is a fit with eq 1 for the α-process, plus the β-process and a conductivity term, while the colored lines represent the ILM fit with β- and MWS processes, plus conductivity.

The dielectric loss of the PNCs in Figure 2 is seen to decrease at low frequencies as expected due to MWS and conductivity, before reaching a minimum, followed by the maximum representative of the segmental relaxation of the polymer. This maximum is positioned at the right in the pure matrix. The α-relaxation is then slowed down as soon as the bare NPs are embedded in the polymer matrix (blue curve), before the peak moves back to the right with surface modification. This phenomenological observation is fundamental for this article, and confirms previous findings with a longer graft ($C_{18}$ instead of $C_8$, see [12]): the nanoparticle surface induces a slow-down of the neighboring polymer chains, while modification of the same surface with small silane molecules counteracts this slow-down. We will now analyze the modification of the α-relaxation in terms of the ILM, which describes the total dielectric response in terms of a bulk (with unmodified dynamics with respect to the neat polymer) and an interfacial contribution, as well as their interferences. As a result, the loss and storage permittivity responses are quantitatively and simultaneously described, and the fits describe well the data in Figure 2 (as well as the other data shown in the SI). Details of the fits are displayed in Figure S4 in the SI, where the different contributions – MWS, α (including both IL and bulk contributions which are linked), β, and conductivity – to the dielectric loss are highlighted for two selected samples.

The main results of the dielectric analysis of these nanocomposites are the properties of the interfacial layer. An example of analysis of the interfacial layer is provided in Figure 3a, and the fit parameters are reported for PNCs at all volume fractions in Figure 3b and 3c. In Figure 3a, the dielectric loss of a pure interfacial layer deduced from the fit of experimental data but without bulk contribution ($\Phi_{bulk} = 0$), $\varepsilon_{IL}''(\omega)$, is represented for PNCs with 30%v of silica and different grafting densities as given in the legend. It corresponds to the sum of two HN functions for the α- and β-processes as deduced from the ILM fit.

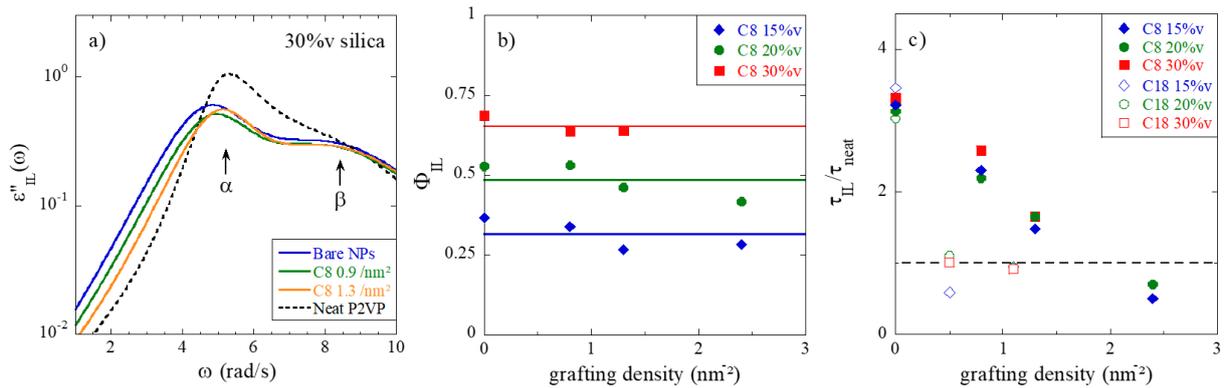

**Figure 3**. **(a)** Dielectric loss of a hypothetical system made of 100% of interfacial layer for P2VP PNCs ($\Phi_{NP}$ = 30%v, T = 423 K, silane grafting as indicated in the legend). **(b)** Volume fraction of interfacial layer for PNC series at 15%v, 20%v, and 30%v volume fraction, as a function of $C_8$-grafting density. **(c)** Segmental relaxation time of the interfacial layer, relative to the neat, at 423 K for the same series as in **b**.

Two important parameters of the ILM analysis are the interfacial layer volume fraction, $\Phi_{IL}$, given here with respect to the polymer part ($\Phi_{IL} + \Phi_{bulk} = 1$), and its characteristic time. The first is plotted in Figure 3b. Two remarkable features can be seen: first, the volume fraction of the interphase is independent of the surface modification. Secondly, it increases with the silica volume fraction, i.e., the available silica surface. $\Phi_{IL}$ can thus be used to determine the thickness of the interfacial layer. A simple cubic model calculation, which ignores silica interactions and thus real particle arrangement gives a thickness of 3.7±0.1 nm. A more accurate model taking into account the particle positions from scattering and simulations and the overlapping volume is discussed below. Whatever the exact value, it is striking to see that the thickness of the interfacial layer is independent of experimental conditions, like silica content and grafting. Therefore, it seems to reflect an intrinsic property of the polymer-surface couple.

Figure 3c, finally, represents the characteristic time of the interfacial layer, expressed as a ratio to the value of the neat polymer. The selected temperature (423 K) corresponds to ca. T = 1.1 $T_g$, which is the closest temperature to $T_g$ where the overall segmental dynamics of P2VP is well-visible in the BDS frequency window. The increase of the ratio $\tau_{IL}/\tau_{neat}$ thus represents the slow-down of the polymer dynamics in the vicinity of the silica (i.e., within ca. 4 nm of the surface, as discussed in the preceding paragraph). It is interesting to see that the PNCs with bare particles possess an interfacial layer with virtually the same relaxation time for all silica contents. As surface modification is introduced, the dynamics accelerates, i.e., the ratio decreases, and seems to level off at around one, which corresponds to unperturbed segmental dynamics. The effect of silane grafting on the slow-down can thus be followed, and above approximately 1.5 nm$^{-2}$, the pure P2VP dynamics is recovered. For comparison, we have superimposed the main result of a previous analysis with a similar but longer

silane molecule, $C_{18}$. [12] The effect on $\tau_{IL}$ is considerably stronger, and with already 0.5 $C_{18}$ per $nm^2$, the pure P2VP relaxation is reached. Although the statistics of the two curves in Figure 3c are insufficient for a precise determination of such a threshold value, it is clear from the decays that the relaxation time of the $C_{18}$-samples has already joined the pure P2VP value at 0.5 $nm^{-2}$, while the $C_8$-PNCs will do the same somewhere between 1.5 and 2 $nm^{-2}$. The ratio between the two thresholds is thus of the order of 3 or 4, which is considerably larger than the ratio of about two between the alkyl chain masses (18:8). As for a given surface grafting density, the total mass of alkyl chains surrounding a given NP is directly related to the molecular weight of the graft, it can be concluded that the effect on the dynamics is not simply related to the amount of $CH_2$ groups, but also to their spatial organization. In particular, one may speculate that the $C_{18}$ groups have a higher propensity to cover homogeneously the silica surface providing a more efficient screening effect from the polymer chains. On the opposite, the shorter $C_8$ might form locally dense regions due to a lower steric hindrance, leaving free silica zones, i.e., covered by hydroxyl groups favoring polymer adsorption. Our findings are in qualitative agreement with recent results from atomistic molecular dynamics simulations of silica-filled polyisoprene, where planar silica substrates were covered with silane of different alkyl lengths ($C_3$ and $C_8$, i.e., with 3 or 8 carbon atoms in the alkyl part) and different grafting densities. [16] It was found that the slowing-down of the polymer dynamics due to polymer adsorption is weakened upon silane grafting with a stronger effect of the longest graft at high grafting density. This effect is concomitant with an increase of the diffusion coefficient of the adsorbed chains, which almost reaches the one of the bulk polymer chains.

The results shown in Figure 3c demonstrate that it is possible to tune the interfacial dynamics using grafts with longer or shorter alkyl chains, at different concentrations. The remaining question to be answered is how such a surface modification affects the structure of the nanocomposites, and TEM (Figure 1) and in particular SAXS are the most suited methods, to be discussed in the next section.

**C. Structure of PNCs studied by SAXS**

The microstructure of all nanocomposites has been investigated by SAXS. Results for the 15 and 20%v series are reported in the SI. They are conceptually very similar to the 30%v-series shown in Figure 4. In Figure 4a, the scattered intensities are plotted for different grafting densities. The intensities are compared to the form factor measured at high dilution as discussed before. At high q, where the intensity is sensitive to the surface and the shape of the particle, a good superposition is observed. Around $3.3 \times 10^{-2}$ $Å^{-1}$, the PNC intensities at these high concentrations begin to deviate. For bare NPs, or low grafting density up to 1.3 $nm^{-2}$, the curves present a well-defined peak around $2.8 \times 10^{-2}$ $Å^{-1}$. At the highest grafting of 2.9 $nm^{-2}$, the curve shows a completely different spatial organization. As already visible at 2%v (see SI and discussion above), there is a deep correlation hole, and the intensity deviates from the form factor at higher q-vectors. At low q, a strong upturn is found. This low-q increase translates the attractive interactions between nanoparticles, inducing aggregation. They are triggered by the suppression of attractive polymer-silica interactions caused by surface modification, and thus of the steric protection against aggregation. Depletion interactions induced by the polymer chains (which are about a factor of two smaller than the particles) may also participate in generating interparticle attraction. All these features correspond to aggregation and large-scale spatial fluctuations induced by the high grafting density and they correspond to those reported for nanocomposite melts by Hall et al. [47] These authors experimentally varied the interfacial attraction via the polymer. They studied poly(ethylene oxide) and polytetrahydrofuran (PTHF)-systems, the latter being less attractive because of lesser hydrogen bonding with the silica. A decrease of the interfacial attraction in PTHF reduces local order and thus leads to a low-q increase, and a structure factor peak shifted towards higher q. PRISM integral equations describe these features, and provide a satisfactory description of polymer-mediated NP concentration fluctuations. The latter ultimately induce depletion aggregation and microphase separation. In our case, increasing coating coverage decreases the polymer-NP effective attraction [12] with a qualitatively similar behavior as predicted by PRISM in terms of peak shift and low-q upturn.

By comparing the families of curves at 15 and 20%v of silica (see Figure S5 in SI) to the 30%-curves in Figure 4, the series at higher volume fractions are found to bundle at lower intensities at low q. This is the natural consequence of the increase in silica content highlighting more and more the hard-core repulsion between particles. This repulsion induces a decrease in the apparent compressibility. As the surface modification is added, some intermediate upturns at 2.4 $nm^{-2}$ can be seen at 15 and 20%v of silica, showing that grafting affects NP interactions in a progressive (and thus tunable) way. One can also follow the peak positions as a function of volume fraction, for bare NPs and intermediate grafting (while it disappears at the highest grafting): at 15%v, the peaks correspond to center-to-center distances of 27 – 31 nm, which are larger than two particle radii, and indicates that NPs still have quite some space to re-organize. At 20%v, the upturn is more prominent at low q, but the peaks remain well-defined leading to distances of ca. 24 to 26 nm. At 30%v, finally, the peak position corresponds to a center-to-center particle distance of ca. 22.5 nm. This distance expresses the fact that particles interact repulsively due to their hard cores, and they do not have much space to reorganize at this concentration. There does not seem to be any systematic dependence with the amount of surface modification. Indeed, we found that the peak positions follow a $\Phi^{-1/3}$-law for the three series (see SI, Figures S6 and S7) and the slight variations observed upon grafting are compatible with the variation in volume fraction between samples (Table 1).

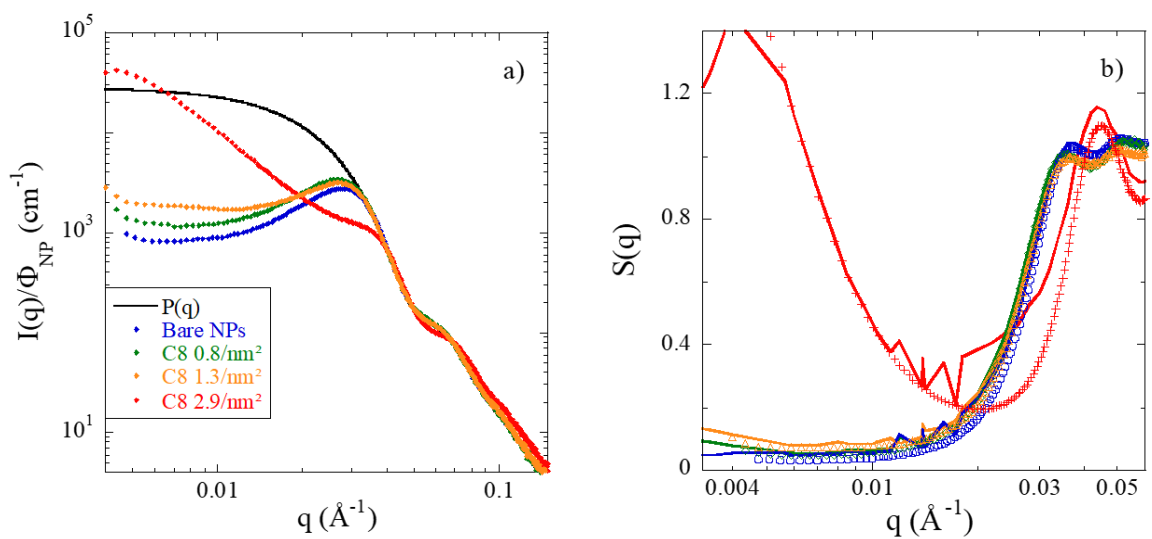

**Figure 4**. **(a)** SAXS results of P2VP PNCs with 30%v of silica: scattered intensity as a function of wave vector, for different grafting densities as given in the legend. **(b)** Apparent structure factors obtained by dividing by the average form factor of the NPs.

In Figure 4b, the apparent structure factors obtained by division of the experimental intensities by the average form factor of the nanoparticles as measured in dilute suspension are presented. The apparent structure factor is a generalization of the structure factor of monodisperse spheres, where it is given by the Fourier transform of the pair-correlation function. Here, all pair correlations are weighted by the different particle volumes, and the apparent structure factor roughly coincides with the true one for low enough polydispersities. In any event, in all model calculations below, the polydispersity is taken into account. The apparent structure factors provide the same information as the intensities, but they focus on the interactions. For instance, absence of interactions as in ideal gases would result in a constant, 1. In this log-lin representation, one observes the close vicinity of all curves besides at the highest grafting. In particular at low q, the highly repulsive character is visible – the low-q limit for monodisperse spheres would give exactly the compressibility relative to an ideal gas. At intermediate q, a peak is reached, with higher order peaks decreasing towards one. Once again, the notable exception is the PNC at the highest grafting, where the structure factor has a completely different shape: the low-q upturn and the correlation hole are clearly visible, and at high q, a strong peak is reached at a position at the right-hand-side of the other samples. This peak position corresponds to a shorter typical center-to-center distance of 14 nm, to be compared to $2R_{NP}$ = 17 nm. The same is observed for the lower volume fractions, 15 and 20%v. Our interpretation of this result is that the system has developed into a highly aggregated one, where the internal structure is probably optimized towards higher density by peculiar correlations between larger and smaller beads.

The apparent structure factors measured for all PNCs at all silica contents have been fitted by a reverse Monte Carlo algorithm, [33-35] as described in the methods section. The result is a sequence of particle configurations the scattering of which is compatible with the measured structure factor. In Figure 4b, the corresponding fit functions have been superimposed to the data. The fit quality is seen to be quite good for all samples but the highest grafting densities, where the strong correlation hole is seen to be difficult to reproduce. The origin of this mismatch is unclear at the current stage, but it might be due to deviations of the real silica NPs from ideal spheres, which may affect local interactions in close contact. Once the sequence of configurations is available, a statistical analysis of the particle positions can be performed. As we are interested in the action of the grafts and possibly adsorbed polymer on the correlations between particles, the interparticle spacing (IPS) distribution function has been deduced from the particle positions. This function is a generalization of the pair-correlation function, for polydisperse spheres, with focus on the interparticle spacing: it expresses the number of times a certain surface-to-surface distance is encountered in the simulation box representing the configuration. The results are plotted for all samples in Figure 5 in terms of the normed IPS, i.e., the number of times a given surface-to-surface distance is found, normed to the same number for a hard-sphere gas of same characteristics. For an analysis of uncertainties in IPS and a comparison with g(r) determined from the particle centers, see SI (Figures S8 and S9).

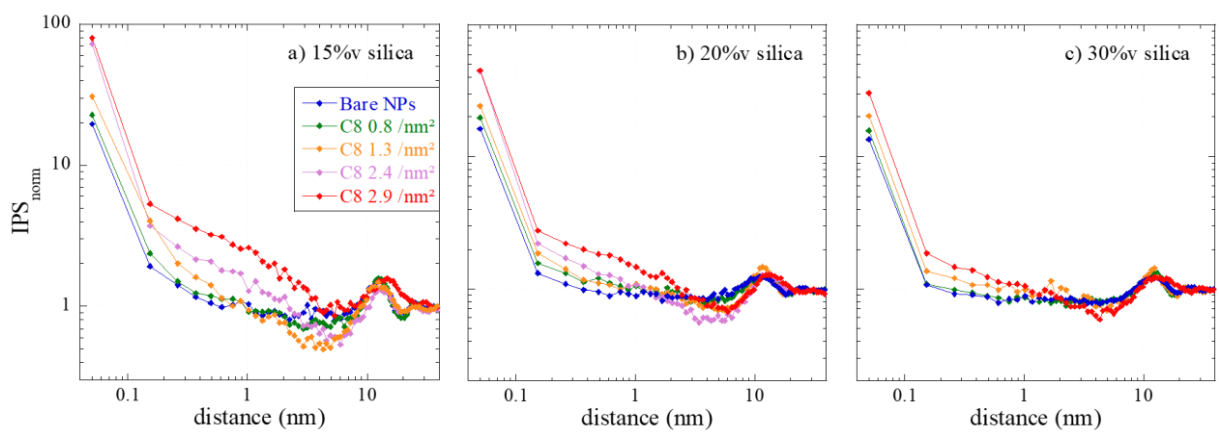

**Figure 5**. Interparticle spacing function (IPS) vs. surface-to-surface distance, normed to the same quantity in a hard-sphere gas of same parameters (concentration, size distribution), for different grafting densities as indicated in the legend. The nominal silica volume fractions of the PNCs are **(a)** 15%v, **(b)** 20%v, and **(c)** 30%v.

The low-distance limit of the normed IPS function in Figure 5 expresses the probability of contact with respect to the hard-sphere gas. Whereas this number is low for the bare NPs (presumably due to the steric buffer action of the adsorbed polymer chains), the increase in contact probability is seen to be highest for the lowest volume fraction, and for the highest grafting fractions. Moreover, there is a notable preference for particles to be close, within typical distances of ca. 1 nm. Around distances corresponding to about one or two particle radii, i.e., above 10 nm, there is some structure visible, which is due to the second layer of neighbors. At high distances, finally, the normed IPS tends to one, implying that there is no difference at large distances between PNC samples and hard-sphere gases.

When comparing the normed IPS functions for the different volume fraction series, it appears that the family of curves become closer with increasing silica content. Simultaneously, the increase of probability of contact becomes smaller, i.e., it approaches the hard-sphere gas. This means that the impact of grafting on structure at high volume fraction is smaller, presumably due to the higher crowding which leaves less space for reorganization, as discussed directly with the structure factors in Figure 4.

There are different ways to analyze the IPS, or to express it in terms of properties of real space configurations. In Figure 6a, the normed contact probability is plotted for all PNC samples. It is seen to increase with grafting, and to decrease with silica volume fraction. The latter effect indicates that at higher silica contents, the particle structures are closer to that of a hard-sphere gas, and less subjected to changes induced by surface modification. As the first neighboring particle is usually in close contact, and as the distance to any particle is expressed by the IPS, one can integrate the raw IPS function to determine the distance where a second particle is encountered. In Figure 6b, this distance is shown, for all silica contents, as a function of grafting density. The distance is seen to be rather large (around 4 – 5 nm) at 15%v, implying a rather loose assembly of particles at low grafting, before it decreases at high grafting, meaning that on average a second particle is "pulled in" into close contact. This effect is seen to decrease for higher volume fractions, where particle assemblies are denser anyhow. Nonetheless, a critical threshold value of about 1.5 $nm^{-2}$ seems to persist. In Figure 6b, there appears to be an artefact at 15%v, where an increase in the second-particle distance is seen at 1.3 $nm^{-2}$. We have attempted to understand the origin of this increase and we have checked if it can be traced back to slight uncertainties in the absolute intensity. We have therefore repeated the same Monte Carlo analysis for three sample data sets at 1.3 $nm^{-2}$, shifting two of them by ± 2% which corresponds to the uncertainty in positioning the sample intensity with respect to the particle form factor (see SI, Figure S8). The resulting indicators in Figure 6, including the "two particle distances", have been converted into an error bar. The increased value at 1.3 $nm^{-2}$ and 15%v of silica in Figure 6b is seen to persist within the error bar. We conclude that while we do not have any physical explanation for such an effect, its origin does not lie in a wrong positioning of the scattering curve in absolute intensity. In the last plot, Figure 6c, a different analysis is proposed. Here a fixed distance, corresponding to the length of two $C_8$ molecules (2L = 2.5 nm) has been used as upper bound for the integral over the raw IPS. The result is the number of neighbors typically encountered up to this distance. This number obviously increases with the volume fraction, but also with the grafting density, and the latter effect is again stronger for the lowest silica fraction. Finally, one may note that Figures 6b and 6c are two sides of the same medal, expressed by different parameters.

The different ways of exploring the IPS proposed in Figure 6 illustrate how the NP dispersion in the nanocomposite changes with volume fraction and grafting density. The volume fraction effect has been discussed several times, and is thought to be mainly a crowding effect which approaches the particle configuration to a hard-sphere one. The surface modification effect is more subtle, and it seems to imply the existence of a threshold of ca. 1.5 $nm^{-2}$. Below this threshold, particle dispersions are only slightly affected by the grafting, whereas above it a complete reorganization, with strong aggregation, is observed.

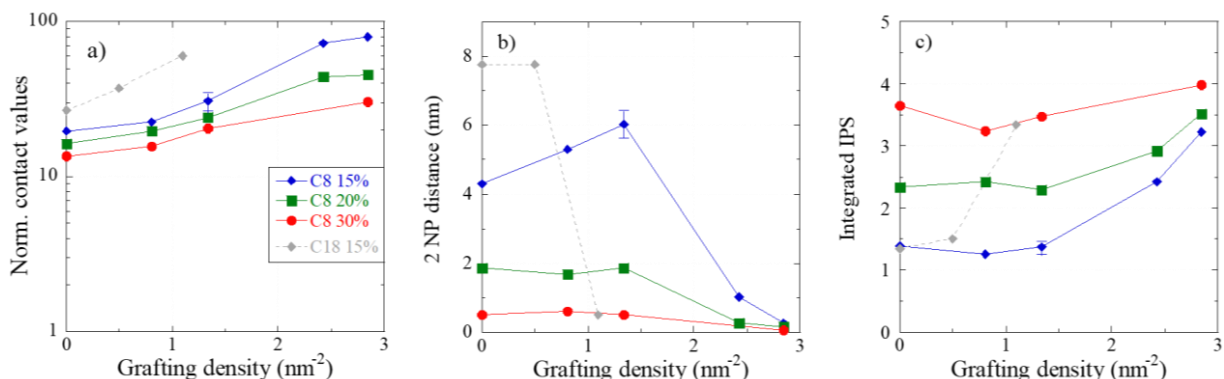

**Figure 6. (a)** Evolution of the normalized contact values with NP volume fraction (15, 20, and 30%v). **(b)** Average second-particle surface-to-surface distances determined by integration of the raw IPS up to a sum of two neighboring NPs. **(c)** Integrated raw IPS from surface to twice the silane length using 2L = 2.5 nm for $C_8$ (resp. 5.1 nm for $C_{18}$). All plots are represented as a function of surface modification, for the three volume fraction series. Results obtained for the 15%v-series with $C_{18}$ surface modification [12] are included in grey for comparison.

It is instructive, finally, to compare the structural indicators obtained with $C_8$ with the corresponding ones for $C_{18}$. For this purpose, we have superimposed in Figure 6 the evolution of the three indicators with the grafting density of $C_{18}$ at 15%v of silica, i.e., when the signature of aggregation with respect to hard spheres is most developed. The contact values are higher for $C_{18}$ than for $C_8$ in Figure 6a, indicating denser assemblies. In parallel, both the strong decrease of the distance between two NPs and the increase of the number of neighbors within a shell corresponding to the size of two silanes are clearly shifted towards lower grafting densities, in Figures 6b and 6c, respectively. As observed for the dynamical features in part B, the influence of the longer alkyl chain length of $C_{18}$ is stronger than $C_8$ to favor NP aggregation by reducing the buffer effect of the polymer segments at the silica surface. As with the dynamics, by again comparing the threshold values in Figure 6, it appears that the $C_{18}$ effect is stronger than the mass effect of (18:8) expected from the ratio between the alkyl chain masses.

### D. Determination of the true interfacial thickness by combining BDS, SAXS, and RMC

A key result of the ILM analysis of segmental dynamics of the interfacial layer discussed above is the volume fraction of the polymer layer slowed-down by the presence of the silica. Based on an idealized cubic model, this volume fraction has been converted into an estimation of the interfacial thickness of a few nanometers. Having measured the particle dispersion by SAXS, and having sets of particle dispersions compatible with this experimental intensity at hand, it is possible to refine this value by taking overlap between particle layers of the obtained configuration explicitly into account. It is thus important to study the relationship between dispersion and interfacial layers more in detail.

The idea is to use the concept of interfacial thickness to characterize the type of particle dispersion. For a given 3D particle arrangement, there should be a specific relationship between the thickness and the volume fraction of the interface, due to overlap. Perfectly ordered and well-dispersed particles, like in a cubic crystal, e.g., have an interfacial volume proportional to the particle surface, as long as the interfacial thickness is small enough to avoid overlap. In presence of overlap, the interfacial volume fraction increases less strongly with thickness than in the ideal case. For the ideal cubic case, simple geometric expressions are available, including overlap. [17] As soon as particles tend to agglomerate, they are, however, of limited use.

In Figure 7a, the evolution of the interfacial volume fraction $\Phi_{IL}^{PNC}$ with a (hypothetical) interfacial layer thickness is plotted exemplarily for different types of dispersion, corresponding to PNC samples with 15%v silica content, with either bare particles or high silane grafting (2.9 nm$^{-2}$). The silica content is also represented and is seen to meet the experimental volume fraction with high accuracy. Note that the silica and the layer volume fractions are determined by the same algorithm based on the positions of N particles in the simulation box, thereby providing a cross-check of the algorithm. Another verification lies in the fact that $\Phi_{IL}^{PNC}$-curves saturate at values approaching 1 - $\Phi_{NP}$, which is why we have left the silica in the definition of $\Phi_{IL}^{PNC}$ ($\Phi_{IL}^{PNC}+\Phi_{NP}+\Phi_{bulk}$ = 1) as opposed to the pure polymer part discussed in Figure 3b.

The two $\Phi_{IL}^{PNC}$-curves in Figure 7a follow different laws despite their close silica contents. For very small thicknesses, PNC with bare NPs display a steeper slope, meaning that the dispersion is better, and more interfacial volume is created with every Angstrom of thickness around the more individually dispersed particles. On the contrary, in the highly aggregated case, there is immediate overlap of interfacial layers, leading to a reduction of volume of the latter. For bare NPs, the maximum available polymer volume is thus reached with ca. 15 nm thickness, while 25 nm are needed to cover all the particle-free regions of the sample in the aggregated case. The evolution of the interfacial volume fraction with thickness shown in Figure 7a for different grafting densities thus characterizes the quality of the dispersion. In Figure 7a, the determination of the real thickness corresponding to the total interfacial layer volume fraction is exemplarily shown, and a thickness of 4.6 nm is found for the bare system, based on an interfacial layer volume fraction determined by BDS of $\Phi_{IL}^{PNC}$ = 0.31.

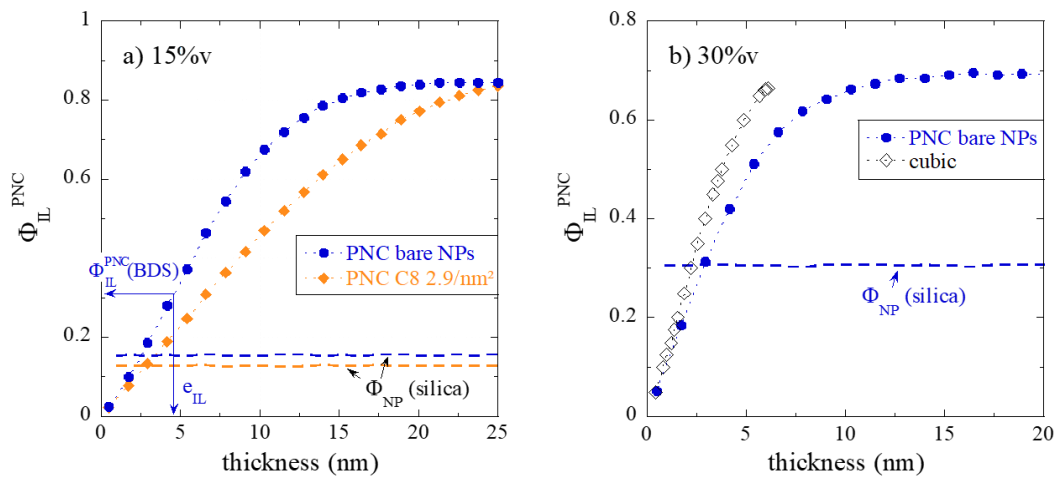

**Figure 7.** Volume fraction of interfacial layer $\Phi_{IL}^{PNC}$ as a function of interfacial layer thickness with respect to the entire sample ($\Phi_{IL}^{PNC} + \Phi_{NP} + \Phi_{bulk}$ =1), for different NP dispersions. The dashed lines represent the silica volume fraction of each sample as determined by the same algorithm. **(a)** 15% PNCs, circles are bare NPs ($\Phi_{NP}$ = 15.3%v), and diamonds are high silane grafting (C$_8$ 2.9 /nm$^2$, $\Phi_{NP}$ = 12.6%v). **(b)** 30% PNCs with bare NPs (30.7%v) compared to the prediction of an (here inappropriate) cubic model with overlap correction. [17]

In Figure 7b, the total interfacial layer volume fraction has been reported for the 30%v-PNC made with bare NPs. Results for other grafting densities are given in the SI. It is found again that the grafting has only limited impact on the dispersion at such high densities, and thus the $\Phi_{IL}^{PNC}$-curves are quite similar. It is instructive, however, to superimpose the prediction of the cubic model (including possible overlap between neighboring layers) used in the literature to the RMC-analysis of experimental data in Figure 7b. Clearly, the large interparticle distance between all spheres on the cubic model leads to a much stronger increase of the $\Phi_{IL}^{PNC}$–function with thickness. It is concluded that it is by no means suitable for dense and possibly aggregated structures as studied here.

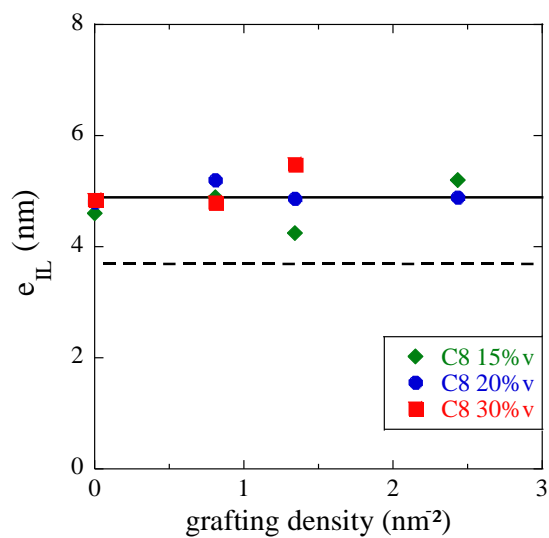

**Figure 8.** Thickness of the interfacial layer determined by a combination of BDS, SAXS, and RMC, as a function of $C_8$ grafting density. The average value over all volume fractions is represented by a solid line. Dashed line: average value considering a cubic NP arrangement with overlap.

The procedure of comparing $\Phi_{IL}^{PNC}$ to the experimental value of the interfacial volume fraction determined by BDS can be generalized to all samples, and the corresponding thickness can be read off following the downward arrow. The interfacial layer thicknesses reported in Figure 8 for all $C_8$-modified nanocomposite samples, as a function of grafting density, represent a key result of the present study, together with Figure 3c where the time scale of the interfacial dynamics is plotted. It needs to be emphasized that the true thickness could only be obtained by a combination of static structural (SAXS) and dynamic methods (BDS), with the help of RMC simulations. As a result, the interfacial layer thickness is found to be remarkably constant with both the grafting density and the silica fraction in PNC. An average value of 4.9±0.2 nm is found, where the error bar has been determined from the standard deviation and the number of points. Moreover, it is found to be compatible to the value obtained with $C_{18}$-surface modification (5.0±0.5 nm) with a lower error bar due to a lesser dispersion of the points [12]. In presence or absence of surface modification of any type ($C_8$ or $C_{18}$), the range of interactions between the polymer segments and the silica surface is thus constant, intrinsic to the polymer-surface couple, or possibly intrinsic to the transmission of cage constraints [13] from one polymer layer to the next. Depending on the grafting, however, the strength of the interaction varies, inducing a stronger (in the case of bare) or weaker (for grafted, longer molecules having a higher impact) slow-down of the segmental dynamics.

## 4. Conclusions

The segmental dynamics of polymer nanocomposites has been studied by BDS, using the interfacial layer model to separate the interfacial polymer volume fraction from its characteristic time. The latter is found to evolve with NP surface modification by $C_8$-grafts: the presence of surfaces of bare silica NPs has a strong impact on the segmental dynamics, resulting in several times longer relaxation times due to the possibility of hydrogen bonding between the pyridine ring of P2VP and the hydroxyl groups at the silica surface. Upon grafting, the interface appears to be screened from the polymer, and the interfacial slowdown of segmental relaxation vanishes. For $C_8$-grafts, this happens above a typical grafting density of about 1.5 nm$^{-2}$, whereas it happens much earlier (ca. 0.5 nm$^{-2}$) for $C_{18}$ grafts. This effect is considerably stronger than the difference in alkyl mass might suggest. Another remarkable result is that the volume fraction of the interfacial layer stays approximately constant for all grafting densities and the graft length.

The dispersion state of the same samples has been studied by SAXS, and the resulting intensity curves interpreted using a reverse Monte Carlo simulation of polydisperse particles in a simulation box. From visual inspection of the apparent NP structure factors, the structure of the nanocomposites is found to depend more strongly on grafting for the lower NP volume fractions than for the higher ones, where less space for rearrangements is available. In any case, at high grafting density, an aggregated state is reached for all silica contents. This interpretation is confirmed by a recently developed analysis, which focuses on the interparticle spacing distribution functions for each sample. These functions can be analyzed in terms of, e.g., probabilities of close contact, or number of neighboring particles in a given shell, and these indicators reflect the morphological transitions triggered by grafting. This SAXS analysis reveals a threshold in grafting density of about 1.5 nm$^{-2}$. Moreover, the three-dimensional representations of the dispersion state available through the RMC simulations allow for a precise determination of the interfacial layer thickness, in agreement with the corresponding volume fraction measured by BDS. It is thus the original combination of BDS, SAXS, and RMC which enables the determination of the nanometric interfacial layer thickness. Finally, the evolution of the interfacial layer volume fraction with (hypothetical) thickness for different dispersions shows that this function is also a valuable tool for the analysis of particle dispersions.

The impact of the alkyl-chain length of the silane graft is found to be stronger for the longer molecules, both on segmental dynamics and on the structure. It is noteworthy that the dynamics is modified on a well-defined range of about 5 nm, whatever the amount or type of graft. On the other hand, the relaxation time depends on these parameters, and different behaviors are observed for dynamics and structure. In BDS, the effect of the longer molecules is considerably stronger than the one expected from the nominal increase in total grafted mass, as one can deduce from the shift in the threshold values. In SAXS, the observed shift is also found to be stronger than the one caused by the grafted mass only. Both in dynamics and structure, grafting longer molecules at the same grafting density has thus a stronger impact.

To summarize, surface-modification can be used to control both the particle dispersion and the slow-down of the polymer interphase in nanocomposites. Structure is controlled by introducing (for bare) or reducing (for grafted NPs) a steric buffer action between particles, possibly concomitant to depletion in the latter case, and dynamics by screening direct interactions of polymer molecules with the silica surface. The modification of the segmental dynamics by the presence of surface-modified NPs has a strong impact on $T_g$ and the overall segmental dynamics of the material, and thus on the mechanical properties of the material. In parallel, modifying the particle dispersion, with aggregation or percolation, also influences the mechanical properties like moduli and resistance to rupture, or, in the case of carbon black fillers, conductivity. We believe that this study, which combines several techniques for a precise determination of interfacial thicknesses, will open the way to new investigations and hopefully control of macroscopic material properties by molecular design of interfacial layer properties.

**Author Contributions:** All authors contributed to the manuscript preparation. All authors have read and agreed to the published version of the manuscript.

**Acknowledgements:** This work was supported by the U.S. Department of Energy, Office of Science, Basic Energy Sciences, Materials Sciences and Engineering Division. A.-C.G. and J.O. are thankful for support by the ANR NANODYN project, Grant ANR-14-CE22-0001-01 of the French Agence Nationale de la Recherche. A.P.S. acknowledges partial support for data analysis and discussions by NSF Polymer program (DMR-1904657).

**Conflicts of Interest:** The authors declare no conflict of interest.**References**

1. Jancar, J.; Douglas, J.F.; Starr, F.W.; Kumar, S.K.; Cassagnau, P.; Lesser, A.J.; Sternstein, S.S.; Buehler, M.J. Current issues in research on structure-property relationships in polymer nanocomposites. *Polymer* **2010**, *51*, 3321-3343, doi:10.1016/j.polymer.2010.04.074.
2. Kumar, S.K.; Benicewicz, B.C.; Vaia, R.A.; Winey, K.I. 50th Anniversary Perspective: Are Polymer Nanocomposites Practical for Applications? *Macromolecules* **2017**, *50*, 714-731, doi:10.1021/acs.macromol.6b02330.
3. Heinrich, G.; Kluppel, M.; Vilgis, T.A. Reinforcement of elastomers. *Curr Opin Solid State Mater Sci* **2002**, *6*, 195-203.
4. Mahtabani, A.; Rytöluoto, I.; Anyszka, R.; He, X.; Saarimäki, E.; Lahti, K.; Paajanen, M.; Dierkes, W.; Blume, A. On the Silica Surface Modification and Its Effect on Charge Trapping and Transport in PP-Based Dielectric Nanocomposites. *ACS Applied Polymer Materials* **2020**, *2*, 3148-3160, doi:10.1021/acsapm.0c00349.
5. Le Strat, D.; Dalmas, F.; Randriamahefa, S.; Jestin, J.; Wintgens, V. Mechanical reinforcement in model elastomer nanocomposites with tuned microstructure and interactions. *Polymer* **2013**, *54*, 1466–1479, doi:http://dx.doi.org/10.1016/j.polymer.2013.01.006.
6. Baeza, G.P.; Genix, A.C.; Degrandcourt, C.; Petitjean, L.; Gummel, J.; Schweins, R.; Couty, M.; Oberdisse, J. Effect of Grafting on Rheology and Structure of a Simplified Industrial Nanocomposite Silica/SBR. *Macromolecules* **2013**, *46*, 6388–6394.
7. Schadler, L.S.; Kumar, S.K.; Benicewicz, B.C.; Lewis, S.L.; Harton, S.E. Designed Interfaces in Polymer Nanocomposites: A Fundamental Viewpoint. *MRS Bulletin* **2007**, *32*, 335-340, doi:10.1557/mrs2007.232.
8. Kumar, S.K.; Jouault, N.; Benicewicz, B.; Neely, T. Nanocomposites with Polymer Grafted Nanoparticles. *Macromolecules* **2013**, *46*, 3199-3214, doi:10.1021/ma4001385.
9. Stockelhuber, K.W.; Svistkov, A.S.; Pelevin, A.G.; Heinrich, G. Impact of Filler Surface Modification on Large Scale Mechanics of Styrene Butadiene/Silica Rubber Composites. *Macromolecules* **2011**, *44*, 4366-4381, doi:10.1021/ma1026077.
10. Musino, D.; Genix, A.-C.; Fayolle, C.; Papon, A.; Guy, L.; Meissner, N.; Kozak, R.; Weda, P.; Bizien, T.; Chaussée, T.; et al. Synergistic Effect of Small Molecules on Large-Scale Structure of Simplified Industrial Nanocomposites. *Macromolecules* **2017**, *50*, 5138-5145, doi:10.1021/acs.macromol.7b00954.
11. Das, S.; Chattopadhyay, S.; Dhanania, S.; Bhowmick, A.K. Improved dispersion and physico-mechanical properties of rubber/silica composites through new silane grafting. *Polymer Engineering & Science* **2020**, *60*, 3115-3134, doi:https://doi.org/10.1002/pen.25541.
12. Genix, A.-C.; Bocharova, V.; Carroll, B.; Dieudonné-George, P.; Chauveau, E.; Sokolov, A.P.; Oberdisse, J. How Tuning Interfaces Impacts Dynamics and Structure of Polymer Nanocomposites Simultaneously. *ACS Applied Materials & Interfaces* **2023**, *15*, 7496–7510, doi:https://doi.org/10.1021/acsami.2c18083.
13. Phan, A.D.; Schweizer, K.S. Influence of Longer Range Transfer of Vapor Interface Modified Caging Constraints on the Spatially Heterogeneous Dynamics of Glass-Forming Liquids. *Macromolecules* **2019**, *52*, 5192-5206, doi:10.1021/acs.macromol.9b00754.17